\documentclass[reprint,unsortedaddress,amsmath,amssymb,aps,prl,showpacs]{revtex4-2}
\usepackage{amsmath}%
\usepackage{amsfonts}%
\usepackage{amssymb}%
\usepackage[cal=cm]{mathalfa}
\usepackage{graphicx}
\usepackage{dcolumn}
\usepackage{bm}
\usepackage{braket}
\usepackage{sidecap}
\usepackage{color}
\usepackage{bbm}
\usepackage{nicefrac}
\usepackage{float}
\usepackage{placeins}
\usepackage[dvipsnames]{xcolor}
\usepackage[utf8]{inputenc}
\usepackage[normalem]{ulem}
\usepackage{nicefrac}
\usepackage{dsfont}
\usepackage{xcolor, soul} 
\usepackage[normalem]{ulem} 
\usepackage{siunitx}
\usepackage[citebordercolor=blue]{hyperref}
\usepackage[utf8]{inputenc}
\usepackage{gensymb}
\usepackage{mathtools}

\begin{document}

\title{Phase Vortex Lattices in Neutron Interferometry}

\author{Niels Geerits$^{1}$}
\email{niels.geerits@tuwien.ac.at}
\author{Hartmut Lemmel$^{1,2}$}
\author{Anna-Sophie Berger$^1$}
\author{Stephan Sponar$^1$}
\email{stephan.sponar@tuwien.ac.at}
\affiliation{%
$^1$Atominstitut, Technische Universit\"at Wien, Stadionallee 2, 1020 Vienna, Austria \\ $^2$Institut Laue-Langevin, 71 Avenue des Martyrs, CS 20156, 38042
Grenoble Cedex 9, France}

\date{\today}
\hyphenpenalty=800\relax
\exhyphenpenalty=800\relax
\sloppy
\setlength{\parindent}{0pt}
\begin{abstract}
    A combination of aluminium prisms inserted into a nested loop interferometer is used to generate a neutron phase vortex lattice with significant extrinsic orbital angular momentum, $\langle L_z \rangle\approx 0.35$, on a length scale of $\approx 220$ $\mu$m, transverse to the propagation direction. Our method is a generalization of recently developed magnetic methods, such that we can exploit the strong nuclear interaction. The stronger potential of these prisms allows for the generation of a tighter lattice. Combined with recent advances in neutron compound optics and split crystal interferometry our method may be applied to the generation of intrinsic neutron orbital angular momentum states. Finally, we assert that, in its current state, our setup is directly applicable to anisotropic ultra small angle neutron scattering.
\end{abstract}

\maketitle
\section{Introduction}
First observed in optics \cite{Allen1992,Enk1994,Terriza2007} Orbital Angular Momentum (OAM) of photons has become ubiquitous in physics. Seeing applications in quantum communications \cite{Gibson2004}, astronomy \cite{Harwit2003} and many other areas \cite{Padget2017}, OAM has also been observed in massive free particles, such as electrons \cite{Uchida2010,McMorran2011}, atoms \cite{Luski2021} and neutrons \cite{Clark2015}. In the latter case, however, there is some ambiguity on whether any of the observed OAM is intrinsic to the neutrons or whether it is simply an extrinsic beam property \cite{Cappelletti2018}, since the dimensions of the phase vortex, henceforth referred to as the vortex diameter, observed in \cite{Clark2015} were much larger than the neutron transverse coherence length (order 100 nm). Furthermore if the neutrons posessed intrinsic OAM the beam could likely only be represented by a mixed state. To address these concerns a method using magnetic prisms was developed to generate a lattice of vortices, with smaller vortex diameters \cite{Sarenac2019}. In addition, the use of static electric fields has been explored \cite{Geerits2021}. However, realistically achievable electric and magnetic potentials are too small to generate vortex diameters on the order of the neutron transverse coherence length. Recently a large lattice of microscopic fork gratings was produced with the goal of producing intrinsic OAM states in a collimated cold neutron beam \cite{Sarenac2022}. It is likely that significant intrinsic OAM was produced. Though important challenges remain such as the production of significant intrinsic OAM in thermal neutron beams or a definitive method of distinguishing intrinsic and extrinsic OAM.
\par
In this paper we generalize the method described by \cite{Sarenac2019}, such that the strong nuclear potential can be exploited, enabling production of smaller vortex diameters at thermal neutron energies. We demonstrate the generation of a vortex lattice using strongly interacting aluminium prisms in a nested loop neutron interferometer. Previously prisms have been employed in neutron interferometry for holography \cite{Sarenac16}. It is expected that neutron OAM will open up new avenues in scattering, allowing one to directly access the complex phase of the scattering amplitude \cite{Afanasev2019,Afanasev2021}. In addition neutron OAM marks an additional degree of freedom, applicable to  quantum information and contextuality \cite{Hasegawa2010,Shen2020}.Coupling OAM to the other degrees of freedom in a neutron has been discussed in other papers such as \cite{Nsofini2016,Sarenac2018}. While not a topic of this paper, in combination with the path, spin and energy of the neutron, the additional degree of freedom, provided by OAM, would enable the first quadruply entangled beam to our knowledge. Quantum mechanical OAM is a type of azimuthal phase structure on the wavefunction of the form $e^{\mathrm{i}\ell\phi}$, with $\phi$ the azimuthal coordinate. Due to continuity conditions $\ell$ can only take on integer values. Therefore OAM, unlike spin angular momentum, is not an intrinsic property of the neutron, but a property arising from the spatial structure of the wavefunction. Hence, OAM states are some times also referred to as spatial helicity states \cite{Ishihara2023}. In an interferometer a single input wavefunction can be split into multiple partial wavefunctions each of which can undergo a simple and independent transformation in each path of the interferometer. When the modified partial wavefunctions of each path are recombined more complicated structures may arise. In a two path, singe loop, interferometer the combination of a phase shifter and a pair orthogonal prisms enables us to generate a composite wavefunction exhibiting azimuthal structure where the $\ell=\pm 1$ mode amplitude is significant. To extract the phase structure of the composite wavefunction an additional reference beam is needed. For this purpose a three path nested loop intereferometer was used.
\section{Results}
\begin{figure}
	\includegraphics[width=8.5cm]{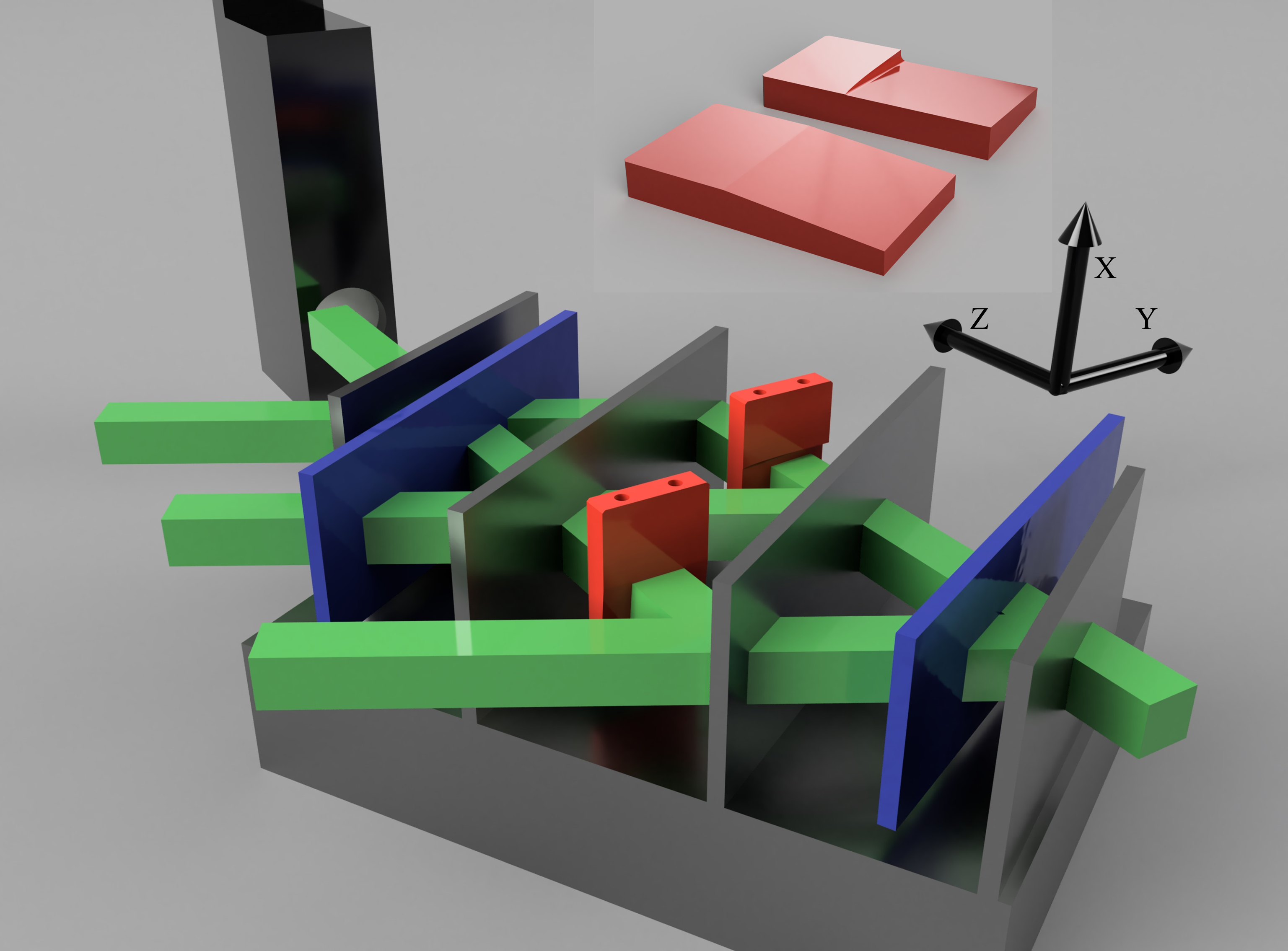}
	\caption{Sketch of the 4 plate interferometer, containing two (red) orthogonal prisms (blown up on the top portion) and two phase shifters (blue). The neutron beam, coming from the right, forms three loops, two small ones between the first and third and second and fourth plate respectively and a large loop between the first and last plate. The phase shifters can be rotated around the vertical to induce phase shifts between the paths in their respective loops. A position sensitive detector is shown in black. Additionally in black the coordinate convention used in this paper is shown.}
	\label{Setup}
\end{figure}

\textbf{Experimental setup.}
The experiment was carried out at a wavelength of 1.92\,\AA \,\,on the thermal neutron interferometry station, S18, at the high-flux reactor of the Institute Laue Langevin (ILL) in Grenoble, France \cite{S18data}. Our setup is shown in figure \ref{Setup}. This interferometer generates three nested loops \cite{Heinrich1988,Hasegawa1996,Filipp2005,Geppert2018}, two small loops between the first and third plate and the second and fourth plate respectively, and a large loop between the first and fourth plate. Our prisms each have a 5 degree slope and are made from aluminium.
To control the phase difference of each loop a minimum of two phase shifters are required. These phase shifters consist of flat silicon and sapphire slabs.\\
\textbf{Theoretical model.} In the case of thermal neutrons where the nuclear potential is low compared to the kinetic energy, the action of a prism can be approximated by a translation of the reciprocal wavefunction (i.e. by convolving with a delta function) $\psi'(\mathbf{k})=\psi_0(\mathbf{k})*\delta(\mathbf{k}-\mathbf{k'})$, while phase shifters imprint a global phase on the wavefunction $\psi'(\mathbf{k})=e^{\mathrm{i}\alpha}\psi_0(\mathbf{k})$. In principle prisms also apply a global phase to the wavefunction, however in this paper we choose to account for this phase in the action of the phase shifter. The input wavefunction (in k-space) is assumed to be Gaussian
\begin{equation}\label{equation:Input}
	\psi_0(\mathbf{k})=\sqrt{\frac{1}{2\pi\zeta^2}}e^{-\frac{(k_x^2+k_y^2)}{4\zeta^2}}\Phi(k_z)
\end{equation}
with $k_x$ and $k_y$ denoting the transverse wavenumbers and $\zeta$ the transverse momentum spread, related to the average divergence of individual neutrons, $\theta$, by $\zeta\approx  k_z\theta$, for small $\theta$. $\Phi(k_z)$ refers to the longitudinal part of the reciprocal wavefunction that is virtually unaffected by the action of the prisms. The composite wavefunction projected from the last interferometer plate to the detector can then be written as
\begin{equation}\label{equation:Composite}
	\psi_1(\mathbf{k})=\frac{1}{\sqrt{3}}[\psi_0(\mathbf{k})+e^{\mathrm{i}\alpha_1}\psi_0(\mathbf{k}-k_\perp\hat{y})+e^{\mathrm{i}\alpha_2}\psi_0(\mathbf{k}-k_\perp\hat{x})]
\end{equation}
where the transverse momentum shift, $k_\perp$, is related to the angle of refraction, $\gamma$, induced by the prisms $k_\perp=k_z\gamma$. It is instructive to look at equation (\ref{equation:Composite}) in real space cylindrical coordinates, $(\rho,\phi,z)$, since the real space equation allows us to more easily deduce the angular momentum properties of this wavefunction.
\begin{equation}\label{equation:RealSpace}
	\psi_1(\mathbf{r})=\frac{1}{\sqrt{3}}\psi_0(\mathbf{r})[1+e^{\mathrm{i}\alpha_1}e^{\mathrm{i}k_\perp \rho\sin(\phi)}+e^{\mathrm{i}\alpha_2}e^{\mathrm{i}k_\perp \rho\cos(\phi)}]
\end{equation}
The expression $\psi_0(\mathbf{r})=\sqrt{\frac{2}{\pi\sigma^2}}e^{-\frac{\rho^2}{\sigma^2}}\Phi(z)$ is the Fourier transform of (\ref{equation:Input}). $\sigma=\frac{1}{\zeta}$ denotes the real space coherence length and $\Phi(z)$ is the real space component of the wavefunction along the $z$ direction. We require that $\Phi(z)$ is normalized (i.e. $\int \mathrm{d}z |\Phi(z)|^2=1$). From here on out it is important to distinguish between the constant reference wavefunction, $\psi_0(\mathbf{r})$, and the test wavefunction $\psi_t(\mathbf{r})$, which is postulated to carry OAM. 
\begin{equation}\label{equation:testwave}
    \begin{aligned}
            &\psi_t(\mathbf{r})=\frac{1}{\sqrt{2}}\psi_0(\mathbf{r})(e^{\mathrm{i}k_\perp \rho\sin(\phi)}+e^{\mathrm{i}\Delta\alpha}e^{\mathrm{i}k_\perp \rho\cos(\phi)}) \\
            &\psi_1(\mathbf{r})=\frac{1}{\sqrt{3}}[\psi_0(\mathbf{r})+\sqrt{2}e^{\mathrm{i}\alpha_1}\psi_t(\mathbf{r})]
    \end{aligned}
\end{equation}
with $\Delta\alpha=\alpha_2-\alpha_1$. We note that the above wavefunctions are not properly normalized, since they do not represent the total neutron wavefunction emerging from the interferometer, but only the part of the wavefunction projected towards the detector.

\begin{figure*}
	\includegraphics[width=17cm]{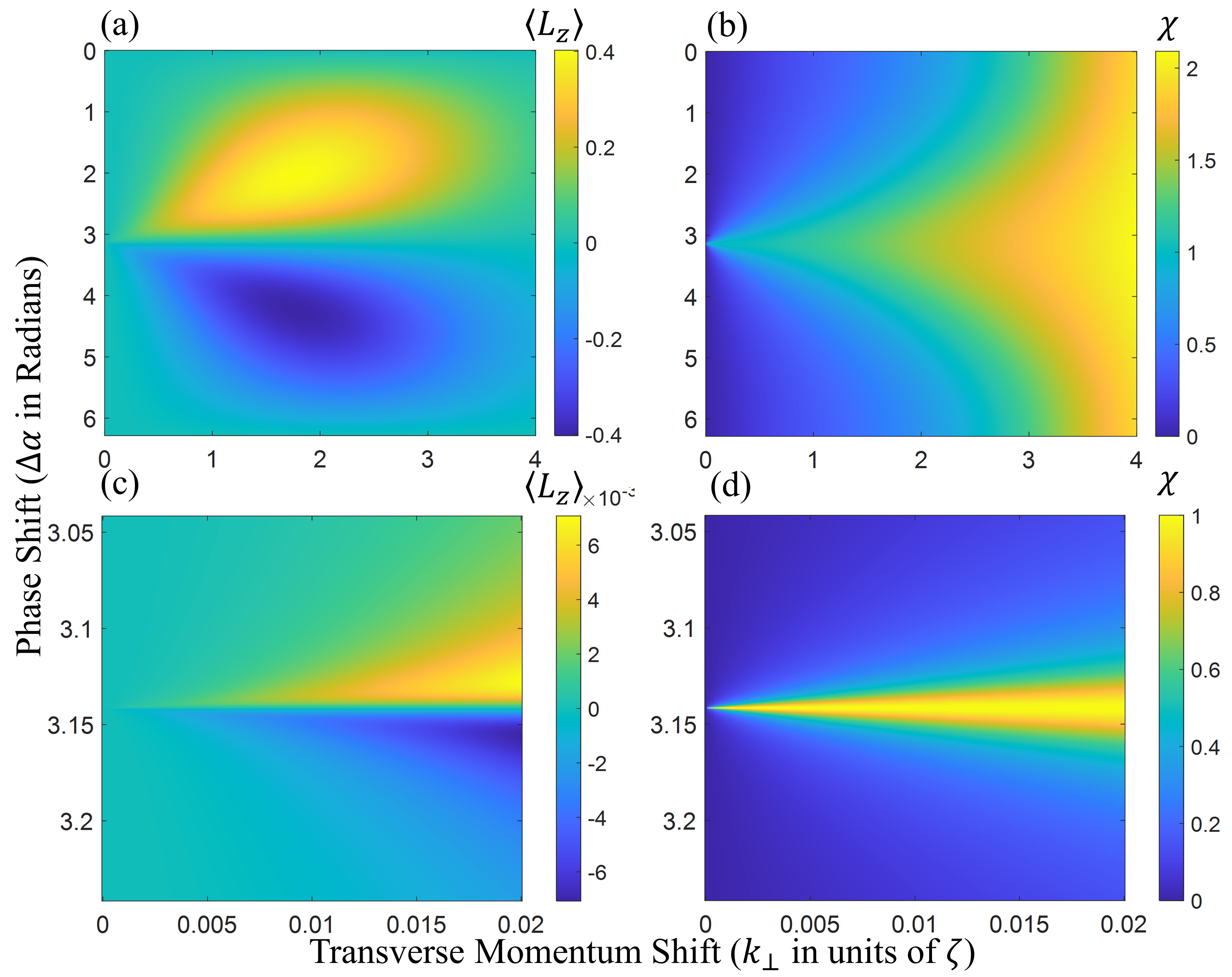}
\caption{(a) Expectation value of the test wavefunction (eq. \ref{equation:testwave}) as given by the analytical expression in equation \ref{equation:TotalOAM} for various transverse momentum shifts $k_\perp$ and phase shifts $\Delta\alpha$. Around $\Delta\alpha=\pm\alpha/2$ and $k_\perp=4\pi$ the OAM attains a maximal/minimal value of $\pm 0.4$ (b) The OAM bandwidth defined by equation \ref{equation:OAMVAR} for $\psi_t$ as a function of transverse momentum shift $k_\perp$ and phase shift $\Delta\alpha$. Inserts (c) and (d) show the behavior of $<L_z>$ and $\chi$ respectively for small $k_\perp$ in the vicinity of $\Delta\alpha=\pi$. In all figures $k_\perp$ is in units of $\zeta$. In the case of the described experiment the normalized $k_\perp$ ranges from $10^{-5}$ (vertical refraction) to $0.02$ (horizontal refraction).} \label{ExpectationBandwidth}
\end{figure*}

\textbf{Treatment of OAM.}
To calculate the total OAM of a wavefunction around the z-axis (propagation direction) we introduce the OAM operator
\begin{equation}\label{OAMOperator}
    L_z=-\mathrm{i}[x\frac{\partial}{\partial y}-y\frac{\partial}{\partial x}]=-\mathrm{i}\frac{\partial}{\partial \phi}
\end{equation}
and its expectation value \begin{equation}\label{equation:Expectation}
    <L_z>=-\mathrm{i}\frac{\int \mathrm{d}\mathbf{r} \psi^*(\mathbf{r})\frac{\partial}{\partial \phi}\psi(\mathbf{r})}{\int \mathrm{d}\mathbf{r} |\psi(\mathbf{r})|^2}
\end{equation}
Applying this calculation to the test wave function described in equation \ref{equation:testwave} we can determine the expected average OAM for various combinations of transverse coherence lengths and refraction angles. It can be shown (see Appendix) that integrating equation \ref{equation:Expectation} over $\phi$ for $\psi_t(\mathbf{r})$ leads to the following expression for the total OAM
\begin{equation}
    <L_z>=\sqrt{2}\pi\sin(\Delta\alpha)\frac{\int \mathrm{d}\rho k_\perp\rho^2|\psi_0|^2J_1(\sqrt{2}k_\perp\rho)}{\int \mathrm{d}\mathbf{r} |\psi(\mathbf{r})|^2}
\end{equation}
In the case of our Gaussian $\psi_0(\mathbf{r})$ this is a standard Hankel transform with the result
\begin{equation}\label{equation:TotalOAM}
    <L_z>=\sin(\Delta\alpha)\frac{k_\perp^2\sigma^2}{4 N}e^{-\frac{k_\perp^2\sigma^2}{4}}
\end{equation}
With N the normalization parameter $N=\int \mathrm{d}\mathbf{r} |\psi(\mathbf{r})|^2=1+\cos(\Delta\alpha)e^{-\frac{k_\perp^2\sigma^2}{4}}$
For large $k_\perp$, the normalization parameter goes to unity. We can easily see in this limit the OAM is maximal/minimal for $\Delta\alpha=\pm\pi/2$. In addition using $N\approx 1$ and the derivative of equation \ref{equation:TotalOAM} we find the approximate value of $k_\perp$ for which the OAM is maximized/minimized: $k_\perp=\pm \frac{2}{\sigma}=\pm 2\zeta$. That is to say that the refraction angle must be about one order of magnitude larger than the average momentum spread of an individual neutron for maximal OAM. Another interesting region of equation \ref{equation:TotalOAM}, is found for small $k_\perp$ in the vicinity of $\Delta\alpha\approx \pi$. Here, around $\Delta\alpha=\pi$, the OAM may vary rapidly and even attain a significant value for a relatively small value of $k_\perp$. \\
The form of equation \ref{OAMOperator} seems to imply that the total OAM depends on the choice of the location of the z-axis in the x-y plane. However for some wavefunctions $<L_z>$ is translation invariant. In these cases the OAM is intrinsic \cite{Berry1998,Neil2002}. For particles it can be shown that under a translation (with $x'=x-x_0$ and $y'=y-y_0$) the OAM changes by
\begin{equation}\label{equation:DLZ}
    <\Delta L_z>=-\mathrm{i}\int \mathrm{d}\mathbf{r} \psi^*(\mathbf{r})[x_0\frac{\partial}{\partial y}-y_0\frac{\partial}{\partial x}]\psi(\mathbf{r})
\end{equation}
Thus it follows that OAM is intrinsic if the expectation values of both transverse momentum components are zero.
\begin{equation}\label{equation:intrinsic_condition}
    <k_x>=<k_y>=0
\end{equation}
Since for our setup $<k_x>=<k_y>=k_\perp$ and $k_\perp\sigma$ is at most $0.01$ we can consider the OAM to be quasi intrinsic, since $k_\perp r_0 \approx 0$. As the interaction range of the neutron is proportional to its' coherence length it does not make sense to look at $r_0>>\sigma$ when examining the OAM of single neutrons.
In addition to the expectation value it is instructive to look at the OAM spread, defined as a standard deviation: 
\begin{equation}\label{equation:OAMVAR}
    \chi=\sqrt{<L_z^2>-<L_z>^2}
\end{equation}
with the second moment given by (see appendix for a complete derivation)
\begin{equation}\label{equation:secondmoment}
    <L_z^2>=\frac{k_\perp^2\sigma^2}{4N}-\cos(\Delta\alpha)\frac{ k_\perp^4\sigma^4}{16N}e^{-\frac{\sigma^2k_\perp^2}{4}}
\end{equation}
It can be seen that the OAM bandwidth is maximal for a phase shift $\Delta\alpha=\pm \pi$. Both the OAM bandwidth and the expectation value are shown for a variety of $\Delta\alpha$ and $k_\perp$ (in units of $\zeta$) in figure \ref{ExpectationBandwidth}. At this point it should be pointed out that in the case of perfect crystal neutron interferometry, the momentum spread $\zeta$ is direction dependent, such that the input wavefunction should be written as
\begin{equation}\label{equation:trueInput}
    \psi_0(\mathbf{k})=\sqrt{\frac{1}{2\pi\zeta_x\zeta_y}}e^{-\frac{(\zeta_y^2k_x^2+\zeta_x^2k_y^2)}{4\zeta_x^2\zeta_y^2}}\Phi(k_z)
\end{equation}
where $\zeta_x$ and $\zeta_y$ differ by three orders of magnitude. Nonetheless the above theory for isotropic momentum spread ($\zeta_x=\zeta_y=\zeta$) is still valid if the transverse momentum shifts induced by the prisms are adapted to the momentum spread in the respective direction. However, the experiment described in this paper employed identical prisms, hence  it is possible that figure \ref{ExpectationBandwidth} does not give an accurate representation of the quasi-intrinsic OAM of our wavefunction. Nonetheless, when we calculate equation \ref{equation:Expectation} analytically, using the $\psi_0(\mathbf{r})$ implied by equation \ref{equation:trueInput}. It can then be shown that
\begin{equation}\label{equation:ASOAM}
    <L_z>=\sin(\Delta\alpha)\frac{k_\perp^2(\sigma_x^2+\sigma_y^2)}{8N}e^{-\frac{k_\perp^2(\sigma_x^2+\sigma_y^2)}{8}}
\end{equation}
\begin{figure*}
	\includegraphics[width=18cm]{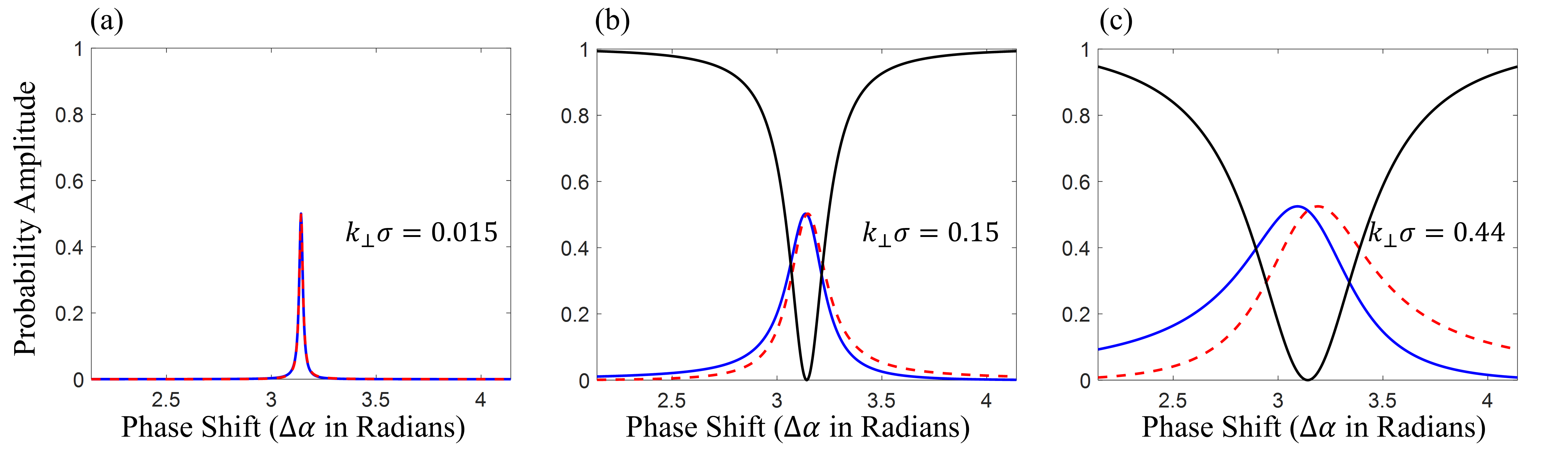}
	\caption{Probability amplitudes of the $\ell=1$ (blue), $\ell=-1$ (red dashed) and $\ell=0$ (black) plotted against the phase shift $\Delta\alpha$ (centered on $\Delta\alpha=\pi$) for various transverse momentum shifts, (a) equal to the experimental case $k_\perp\sigma=0.015$, (b) ten times larger and (c) thirty times larger than in the experimental case. In (a) the $\ell=0$ amplitude is not plotted for improved visibility. It can be clearly discerned that $\ell=\pm 1$ probabilities widen for increasing refraction, $k_\perp$. In addition the $\ell=1$ and $\ell=-1$ probabilities appear to be mirror images of one another (mirrored around $\Delta\alpha=\pi$).}
	\label{Amplitudes}
\end{figure*}
which in form is identical to equation \ref{equation:TotalOAM}, if we define an effective transverse coherence $\sigma^2=(\sigma_x^2+\sigma_y^2)/2$, the normalization parameter $N$ is then unchanged. So figure \ref{ExpectationBandwidth} can also be considered for anisotropic momentum spreads and the maximal amount of OAM generated by this type of setup is not affected by an anisotropic momentum distribution. It follows that in the experiment described here the effective $k_\perp\sigma$ is on the order of $0.015$  \\
Given this effective $k_\perp\sigma$, we may regard the OAM as quasi intrinsic. However as can be seen in figure \ref{ExpectationBandwidth} the OAM production is small for this configuration.
Nonetheless it is instructive to look at the amplitudes of the wavefunctions first OAM modes for small $k_\perp\sigma$. To this end we introduce the azimuthal Fourier transform (AFT)
\begin{equation}\label{equation:AFT}
	\psi^\ell(\rho,z)=\frac{1}{\sqrt{2\pi}}\int_0^{2\pi}\psi(\mathbf{r}) e^{\mathrm{i}\ell\phi}\mathrm{d}\phi
\end{equation}
and its inverse
\begin{equation}\label{equation:IAFT}
    \psi(\mathbf{r})=\frac{1}{\sqrt{2\pi}}\sum_\ell \psi^\ell(\rho,z)e^{-\mathrm{i}\ell\phi}
\end{equation}
where the probability amplitude of the $\ell$th OAM mode is given by 
\begin{equation}\label{equation:AmplitudeProb}
    A^\ell=\frac{\int \mathrm{d}\rho dz\rho|\psi^{\ell}(\rho,z)|^2}{\sum_\ell \int \mathrm{d}\rho dz  \rho|\psi^{\ell}(\rho,z)|^2}
\end{equation}
Hence by applying the AFT to a wavefunction we can determine the amplitude of each OAM mode individually.
The AFT of the test wavefunction (equation \ref{equation:testwave}) is given by the Jacobi-Anger expansion \cite{Abramowitz}
\begin{equation}\label{equation:Expanded}
    \psi_t^\ell(\rho)=(-1)^{\ell}\frac{2}{\sigma}e^{-\frac{\rho^2}{\sigma^2}}J_\ell(k_\perp \rho)\big(1+\mathrm{i}^{-\ell}e^{\mathrm{i}\Delta\alpha}\big)
\end{equation}
Note that we have dropped the longitudinal part of the wavefunction, $\Phi(z)$ for this analysis.
Realistically the refraction angle induced by a neutron optical prism is much smaller than the beam divergence, therefore the width of the Gaussian envelope in (\ref{equation:Expanded}) is much smaller than the period of the Bessel functions, $J_\ell(k_\perp \rho)$. This implies that linearizing the Bessel functions will yield a good approximation of the OAM amplitudes. We note that in the linear limit only Bessel functions of modes $\ell=-1$, $\ell=0$ and $\ell=1$ are non zero, therefore only these OAM modes play a non-vanishing role in our wavefunction. The approximation yields
\begin{equation}\label{equation:FirstOrder}
\begin{aligned}
	&\psi_t^{\ell=0}(\rho)\approx\frac{2}{\sigma}e^{-\frac{\rho^2}{\sigma^2}}(1+e^{\mathrm{i}\Delta\alpha}) \\
	&\psi_t^{\ell=\pm 1}(\rho)\approx\mp \frac{k_\perp \rho}{\sigma}e^{-\frac{\rho^2}{\sigma^2}}(1\mp \mathrm{i}e^{\mathrm{i}\Delta\alpha})
\end{aligned}
\end{equation}
As previously shown the average OAM $\langle L_z\rangle$ is zero for $\Delta\alpha=\pm \pi$. However this new analysis shows that, despite this, the intrinsic neutron OAM is dominated by an equal superposition of $\ell=\pm 1$ modes, while the $\ell=0$ mode is totally suppressed. We may calculate the probability amplitudes of the $\ell=0$ and $\ell\pm1$ modes, according to equation \ref{equation:AmplitudeProb} using our approximate expressions in equation \ref{equation:FirstOrder}. Figure \ref{Amplitudes} shows these probability amplitudes for various $k_\perp$ around $\Delta\alpha=\pi$. It can be seen that for increasing $k_\perp$ the $\ell=1$ and $\ell=-1$ probabilities widen and begin to separate from one another. It can also be seen that the $\ell=1$ and $\ell=-1$ amplitudes are asymmetric around $\Delta\alpha=\pi$, having a steeper slope to one side of the peak compared to the other side. This results in the OAM becoming net positive for $\Delta\alpha<\pi$ and negative for phase shifts above $\pi$.  \\
\textbf{Macroscopic Treatment} Until now we have considered a microscopic treatment where a single wavefunction is centered on the optical axis. Now we turn to the macroscopic treatment where we consider an ensemble of quasi-paraxial wavefunctions which make up a beam.  This is also the scale at which vortex lattices can appear, which carry macroscopic beam OAM. Since the individual neutrons that make up the beam can be far off-axis, compared to their coherence length, equation \ref{equation:DLZ} predicts that most neutrons will have extrinsic OAM with respect to the axis around which (beam) OAM is defined. In our and most other neutron experiments with OAM the main qualitative difference between extrinsic and intrinsic OAM, that can be observed, is the vortex diameter as can be grasped by looking at equation \ref{equation:DLZ}. As the vortex diameter grows the more variability is introduced to the observed OAM of an individual neutron some distance from the vortex center (assuming $<k_x>\neq0$ and/or $<k_y>\neq0$). As a rough definition we may say in the case of (quasi) intrinsic OAM the vortex should manifest on a length scale comparable to the transverse coherence length, while in the case of extrinsic OAM the vortex may exceed this length by many orders of magnitude. It has been predicted that neutrons carrying intrinsic OAM may interact differently with matter, such as in scattering from microsopic targets \cite{Afanasev2019,Afanasev2021} or polarized nuclear targets \cite{Jach2022}. Hence some neutron scattering and transmission measurements may be able to distinguish between intrinsic and extrinsic OAM.
\\
It is well known that a prism inserted into a single loop interferometer generates a Moire fringe pattern along the refraction direction \cite{Sarenac2018b}. In our nested loop interferometer the Moire patterns generated by each loop are overlaid, thereby creating a lattice like structure. The spatial intensity profile can be calculated using the wavefunction projected to the detector (equation \ref{equation:RealSpace}). The wave function impinging on the detector at position $\mathbf{r}'$ is simply equation \ref{equation:RealSpace} with the input wavefunction $\psi_0(\mathbf{r})$ translated by $\mathbf{r}'$
\begin{equation}\label{equation:ApproxWavef}
	\psi_1(\mathbf{r}-\mathbf{r}')=\frac{1}{\sqrt{3}}\psi_0(\mathbf{r}-\mathbf{r}')[1+e^{\mathrm{i}\alpha_1}e^{\mathrm{i}k_\perp y}+e^{\mathrm{i}\alpha_2}e^{\mathrm{i}k_\perp x}]
\end{equation}
\begin{figure*}
	\includegraphics[width=18cm]{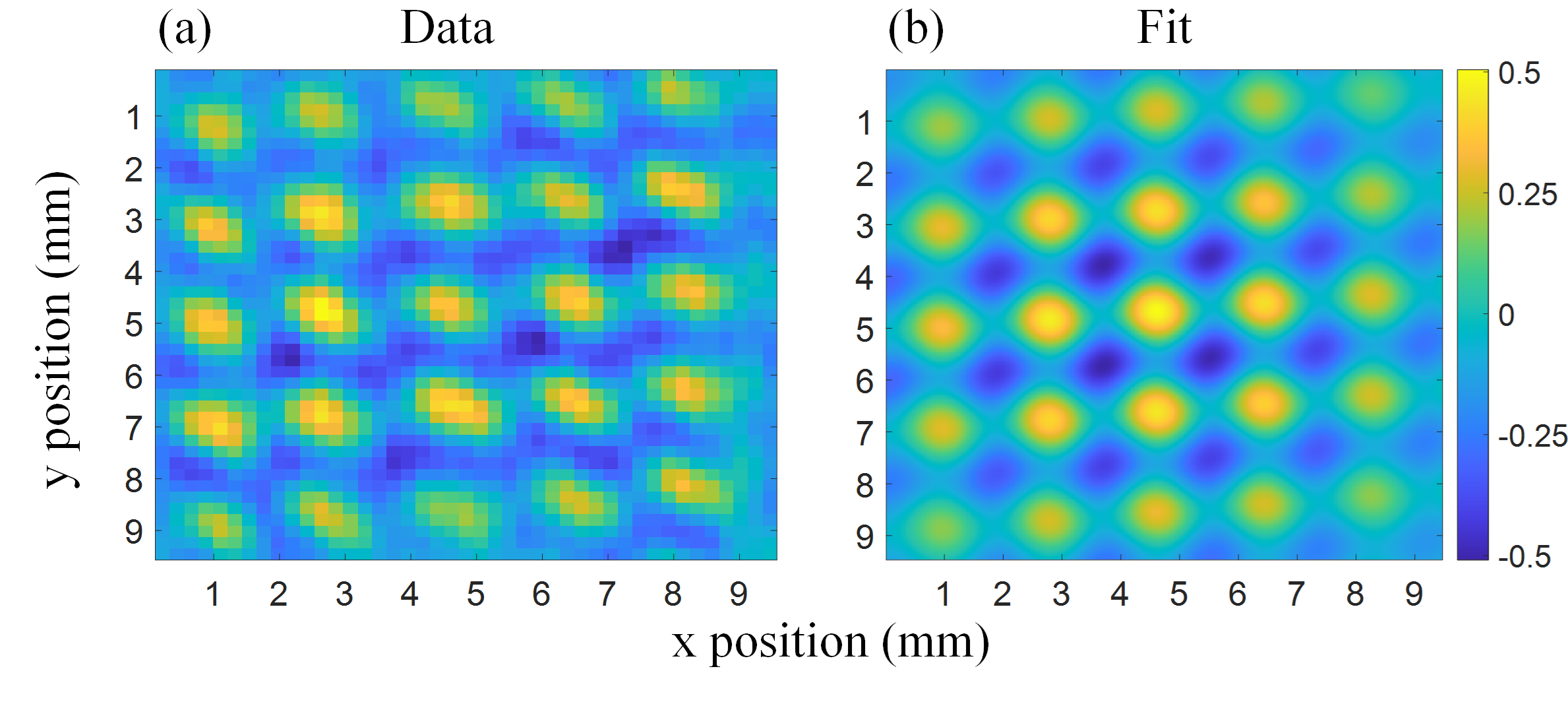}
	\caption{(a) The processed, normalized and filtered image of the neutron vortex lattice, recorded using the position sensitive detector seen in figure \ref{Setup}. The contrast, according to the fit (b) based on (\ref{equation:Intensity}), is 0.53. The lattice period is 1.83 mm.} 
	\label{Results}
\end{figure*}
The intensity profile which is measured can be calculated by taking the absolute value squared of equation (\ref{equation:ApproxWavef}):
\begin{equation}\label{equation:Intensity_Int}
    	I(\mathbf{r'})=\int_\mathcal{P} \mathrm{d}\mathbf{r} |\psi_1(\mathbf{r}-\mathbf{r}')|^2
\end{equation}
with $\mathcal{P}$ a domain given by the pixel size of the detector, which is quasi infinite in size compared to the wavefunction. Assuming $\mathbf{r'}$ falls within the domain $\mathcal{P}$, we may approximate this integral by
\begin{equation}\label{equation:Intensity_Int2}
\begin{aligned}
        	&I(\mathbf{r'})=\frac{1}{3}\int_\mathcal{P} \mathrm{d}\mathbf{r} \delta(\mathbf{r}-\mathbf{r'}) [3+2\cos(k_\perp y +\alpha_1)+ \\ &2\cos(k_\perp x +\alpha_2)+2\cos(k_\perp (x-y) +\Delta\alpha)]
\end{aligned}
\end{equation}
where we used that $|\psi_0(\mathbf{r}-\mathbf{r}')|^2$ may be approximated by a delta function since the coherence length is very small compared to the period of the cosines.
Hence it follows
\begin{equation}\label{equation:Intensity}
\begin{aligned}
    &I(\mathbf{r'})=\frac{1}{3}[3+2\cos(k_\perp y +\alpha_1)+ \\ &2\cos(k_\perp x +\alpha_2)+2\cos(k_\perp (x-y) +\Delta\alpha)]
\end{aligned}
\end{equation}
For the prisms used in this experiment we expect a value of $k_\perp$ which corresponds to a lattice period of $1.75$ mm.
\\
\textbf{Measurements.}
The vortex lattice generated by our setup is shown in figure \ref{Results} (a).
In addition, the figure contains a fit (fig. \ref{Results} (b)) based on equation (\ref{equation:Intensity}). The discrepancies between the fit and the data, could be explained by different amplitudes of the three Moire patterns in equation \ref{equation:Intensity}. These amplitudes can differ depending on the amount of material each partial wavefunction in the interferometer goes through. If two paths "see" a similar amount of material, the amplitude of the Moire fringes from that loop will be large, while if there is a discrepancy in the amount of material, dephasing may occur, thereby lowering the amplitude of the respective loop. \\ 
Since the model used for our fit assumes that the intensity is given by $|\psi_1(\mathbf{r})|^2$, we may extract a part of the test wavefunction, $\psi_t(\mathbf{r})/\psi_0(\mathbf{r})$, from the data, using our model, yielding the phase data needed to compute the amplitude of each OAM mode and the average OAM normal to any domain. Note that since the reconstructed test wavefunction is given by $\psi_t(\mathbf{r})/\psi_0(\mathbf{r})$, we do not observe any coherence effects, as these are all contained within $\psi_0(\mathbf{r})$. Figure \ref{VortexAFT} (a), shows the real part of the reconstructed test wavefunction zoomed in on a single vortex. To calculate the amplitude of each OAM mode we introduce a spatially averaged AFT
\begin{equation}\label{SAAFT}
    \Bar{\psi}_t^\ell=\int_\mathcal{D} e^{\mathrm{i}\ell\phi(x,y)}\psi_t(x,y) \ \mathrm{d}^2\mathbf{x} 
\end{equation}
with $\phi(x,y)$ defined by the argument between the x and y coordinate (i.e. $\phi=Arg(x+iy)$) and $\mathcal{D}$ an arbitrary two dimensional domain, over which the average mode amplitude is to be determined. From the amplitudes calculated in equation (\ref{SAAFT}) the expectation value of the OAM orthogonal to the domain surface can be determined
\begin{equation}\label{equation:Lz}
    \langle L_z \rangle=\frac{\sum_\ell \ell |\Bar{\psi}_t^\ell|^2}{\sum_\ell |\Bar{\psi}_t^\ell|^2}
\end{equation}
To closely approximate equation (\ref{equation:AFT}) a circular domain is chosen to calculate the amplitudes, $\Bar{\psi}_t^\ell$, given by equation (\ref{SAAFT}). To first order it was shown that the $\ell=\pm 1$ amplitudes increase linearly with $\rho$ (equation (\ref{equation:FirstOrder})), hence a larger domain will see a larger maximal value of the OAM. We will, therefore, choose the maximal domain size on which the first order approximations of the test wavefunction are valid. The first order approximation can be used up to $k_\perp \rho=0.75$ with a maximal relative error of less than $0.1$. 
In our setup this corresponds to a domain size of 0.22 mm. Being much larger than the effective transverse coherence of the beam (roughly $5$ $\mu$m) it follows that the OAM must be considered to be extrinsic. The domain on which the spatially averaged AFT is calculated is indicated in figure \ref{VortexAFT} (a). It can be scanned across the reconstructed test wavefunction, $\psi_t(\mathbf{r})/\psi_0(\mathbf{r})$, to calculate $\langle L_z \rangle$ in each section of the image. This OAM expectation value is shown in figure \ref{VortexAFT} (b). Note the diagonal (45 degree) "lines" of constant OAM in figure \ref{VortexAFT} (b), confirming the prediction made by equation \ref{equation:DLZ}.
\begin{figure*}
	\includegraphics[width=17cm]{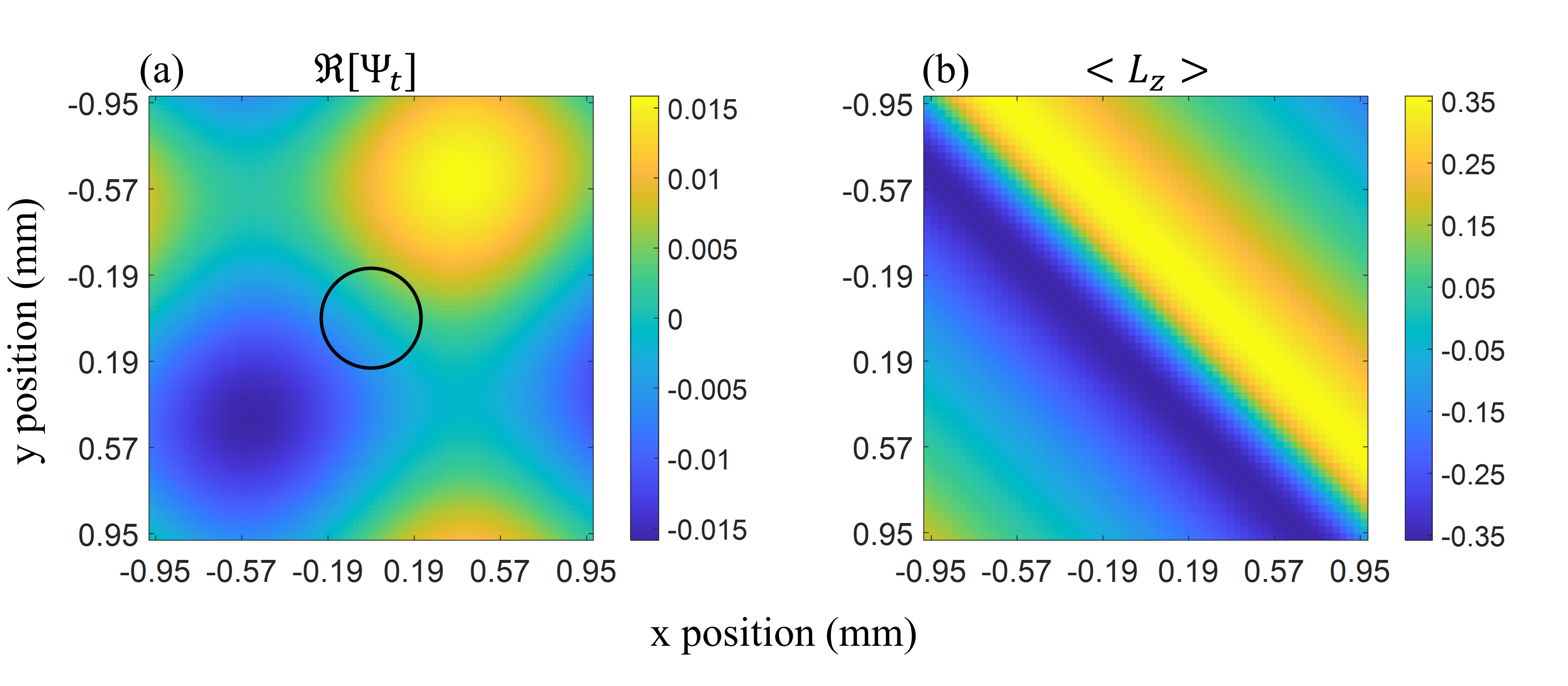}
	\caption{(a) Image of the real part of the macrosopic test wavefunction of a single vortex carrying extrinsic OAM. This test wavefunction is reconstructed using the fit parameters generated by the model shown in figure \ref{Results}. A circle is drawn in the center of the image indicating the domain on which the spatially averaged AFTs are applied and the first order approximations used throughout the paper are valid. The axis around which the OAM is defined is centered on and normal to this ciruclar domain. (b) The average extrinsic OAM $\langle L_z \rangle$ over the image is shown. This is calculated using the spatially averaged AFT (equation \ref{SAAFT}) and equation (\ref{equation:Lz}).}
	\label{VortexAFT}
\end{figure*}
\section{Discussion}
We see that our method using only two prisms generates extrinsic vortices with a significant $\ell=\pm1$ component, such that the average beam OAM can reach up to $|\langle L_z \rangle|\approx0.35$. While the vortex diameter is still much larger than the calculated coherence length and therefore cannot be applied to experiments requiring intrinsic OAM, as described in \cite{Afanasev2019,Afanasev2021,Jach2022}, spatially modulated beams like the one generated in our setup can be applied to ultra small angle scattering. In numerous configurations it has been shown that one dimensional intensity modulation (such as Moire patterns) can be applied to ultra small angle scattering, for example in neutron dark field imaging/Talbot-Lau interferometry \cite{Strobl2008,Strobl2016} and spin echo modulated small angle neutron scattering \cite{Bouwman2009,Bouwman2011,Li2016}. With the exception of a recent development in Talbot-Lau interferometry \cite{Valsecchi2020} the latter methods can only measure the elastic scattering function $S(q)$ in one dimension. Two dimensional intensity modulation, as generated by our setup, could be used to measure two dimensional elastic scattering functions, allowing analysis of anisotropic samples in a single measurement. Such a measurement would employ the same instrument as described in this paper. A sample could be placed between the interferometer and position sensitive detector. Small angle scattering from the sample would wash out the intensity modulation leading to contrast reduction. This contrast reduction is proportional to the Fourier transform of $S(q)$ analogous to \cite{Bouwman2009,Bouwman2011}. By Fourier transforming the modulated intensity pattern it is possible to separate the contrasts of the vertical and horizontal modulation. This allows the instrument to distinguish between vertical and horizontal scattering. Hence, the instrument could simultaneously measure $S(q_x)$ and $S(q_y)$. This scheme could also be applied to the magnetic method for generating vortex lattices \cite{Sarenac2019}. Both the latter method and our approach still lack the focusing prisms used for first order corrections to the divergence/coherence, which prevent dephasing and are available in the one dimensional method \cite{Bouwman2009,Bouwman2011,Li2016}. Though a recent analysis \cite{Thien2023} has demonstrated how to implement first order divergence corrections in a setup analogous to \cite{Sarenac2019} and the setup described in this paper. Focusing elements increase the modulation contrast and allow for larger beam sizes/divergences, thereby increasing the available intensity. These focusing prisms become a requirement when one looks towards generating intrinsic OAM using our method. Equation (\ref{equation:TotalOAM}), shows that the refraction angle of the prisms or $k_\perp$ must be on the same order of magnitude as the beam divergence or $\zeta$, such that the amplitude of the $|\ell|=1$ mode becomes significant. This may be achievable in the near future with recent developments in compound neutron optics \cite{Adachi2002} and micromachining \cite{Kapahi2021}. In addition, steeper prisms made from more dense optical material can be employed in compound devices. The additional space required by obligatory focusing prisms call for larger perfect crystal interferometers. Ongoing developments in neutron interferometry with split crystals may make this possible in the near future \cite{Lemmel2022}. However a fundamental limit is reached as $k_\perp$ approaches the beam divergence $\zeta$ along the diffraction direction, as in this case beams are only poorly diffracted by interferometer plates. For diffraction to efficiently occur the momentum shifted wavefunction $\psi_0(\mathbf{k}-k_\perp\hat{j})$ must have significant overlap with the input wavefunction $\psi_0(\mathbf{k})$, which is defined by the angular acceptance of the interferometer. $\hat{j}$ here refers to the direction normal to the crystal planes. As a result we can estimate that $k_\perp$ can be on the order of $\zeta$. Using equation \ref{equation:TotalOAM} it can be shown that the OAM expectation value cannot exceed $0.1$, due to the diffraction limit. Nonetheless the diffraction limit can be avoided if one uses real space coherent averaging instead of momentum space coherent averaging as was done in this work. That is to say instead of using a composite wavefunction like equation \ref{equation:Composite} where the partial wavefunctions are shifted in momentum space with respect to one another, one could use a composite wavefunction where the partial wavefunctions are shifted in real space relative to each other. 
\begin{equation}\label{equation:FTRST}
    \psi(\mathbf{r})=\frac{1}{\sqrt{2}}[\psi_0(\mathbf{r}-\delta\hat{y})+e^{\mathrm{i}\Delta\alpha}\psi_0(\mathbf{r}-\delta\hat{x})]
\end{equation}
Where real space separations, $\delta$, can be achieved using prism pairs. The Fourier transform of this wavefunction is
\begin{equation}\label{equation:realspacetest}
    \psi(\mathbf{k})=\frac{1}{\sqrt{2}}\psi_0(\mathbf{k})(e^{\mathrm{i}\delta k_\rho\sin(\theta)}+e^{\mathrm{i}\Delta\alpha}e^{\mathrm{i}\delta k_\rho \cos(\theta)})
\end{equation}
with $k_\rho$ denotes the transverse wavenumber, while $\theta$ is the azimuthal angle in momentum space. One can see that this wavefunction is identical in form to the test wavefunction, $\psi_t$ (see equation \ref{equation:testwave}) used throughout this manuscript. Since the OAM operator does not change form under a Fourier transform
\begin{equation}\label{equation:FTLZ}
    \begin{aligned}
        -i[x\frac{\partial}{\partial y}-y\frac{\partial}{\partial x}]&\xLeftrightarrow{\mathcal{F}}-i[k_x\frac{\partial}{\partial k_y}-k_y\frac{\partial}{\partial k_x}] \\
        -i\frac{\partial}{\partial \phi}&\xLeftrightarrow{\mathcal{F}}-i\frac{\partial}{\partial\theta}
    \end{aligned}
\end{equation}
it follows that the OAM of equation \ref{equation:realspacetest} can be derived identically to that of equation \ref{equation:testwave}, detailed in the \textit{Treatment of OAM} section. Therefore the form of the OAM expectation value is identical to that which is described in \ref{equation:TotalOAM}. Contrary to what one may intuitively think the wavefunction in equation \ref{equation:realspacetest} does not obey $<k_x>=<k_y>=0$ it follows that, the OAM is therefore not invariant under translation (see equation \ref{equation:DLZ}) and is therefore also not truly intrinsic. Though once again in some cases for translations within the coherence length, the OAM may be considered quasi-intrinsic. Moving forward in the pursuit of neutron OAM real space coherent averaging methods should be applied since it is technically simpler to generate large real space displacements, $\delta$ of the wavefunction, on the order of the neutron coherence $\sigma$, compared to generating large $k_\perp$ on the order of the wavefunctions momentum spread $\zeta$.
In summary we argue that this work denotes an important step towards high yield OAM generating optical devices for thermal neutrons. Such devices will enable new scattering experiments which can access phase information of the scattering cross section \cite{Afanasev2019,Afanasev2021}. Futhermore, the additional degree of freedom provided by the OAM quantum number would allow quadruple state entanglement in neutrons (energy, position, spin and OAM), opening up new possibilities in the realm of quantum information and contextuality \cite{Hasegawa2010,Shen2020}. In addition, our theoretical analysis, which gives a condition for intrinsic particle OAM (adapted from \cite{Neil2002}), provides a method for determining the probability amplitudes of each OAM mode and examines two special cases of coherent averaging in real and reciprocal space, may be useful in the design of future OAM generating neutron optical instrumentation. Especially the final analysis described in equations \ref{equation:realspacetest}-\ref{equation:FTLZ} could greatly simplify intrinsic neutron OAM production by coherent averaging methods. Finally, we argued that the instrumentation in its current state could be applied to anisotropic ultra small angle neutron scattering, by observing the change in modulation contrast when a sample is placed between interferometer and detector.
\section{Methods}
\subsection{Interferometry Setup}
\begin{figure*}
	\includegraphics[width=17cm]{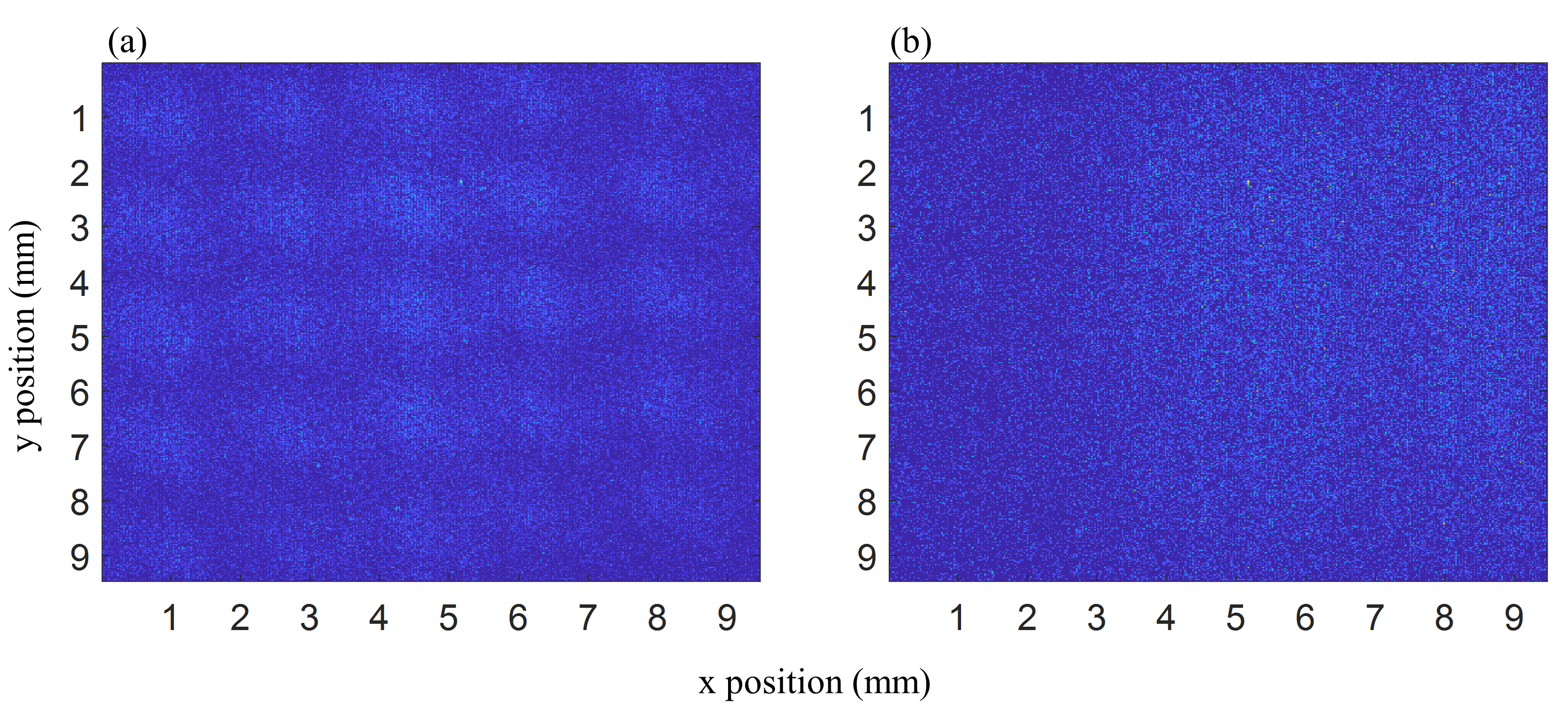}
	\caption{Sum over the raw datasets used to generate the figures shown in this paper. (a) Image with the prisms inserted. (b) Image of the intensity distribution without prisms in the interferometer.}
	\label{raw}
\end{figure*}
\begin{figure*}
	\includegraphics[width=17cm]{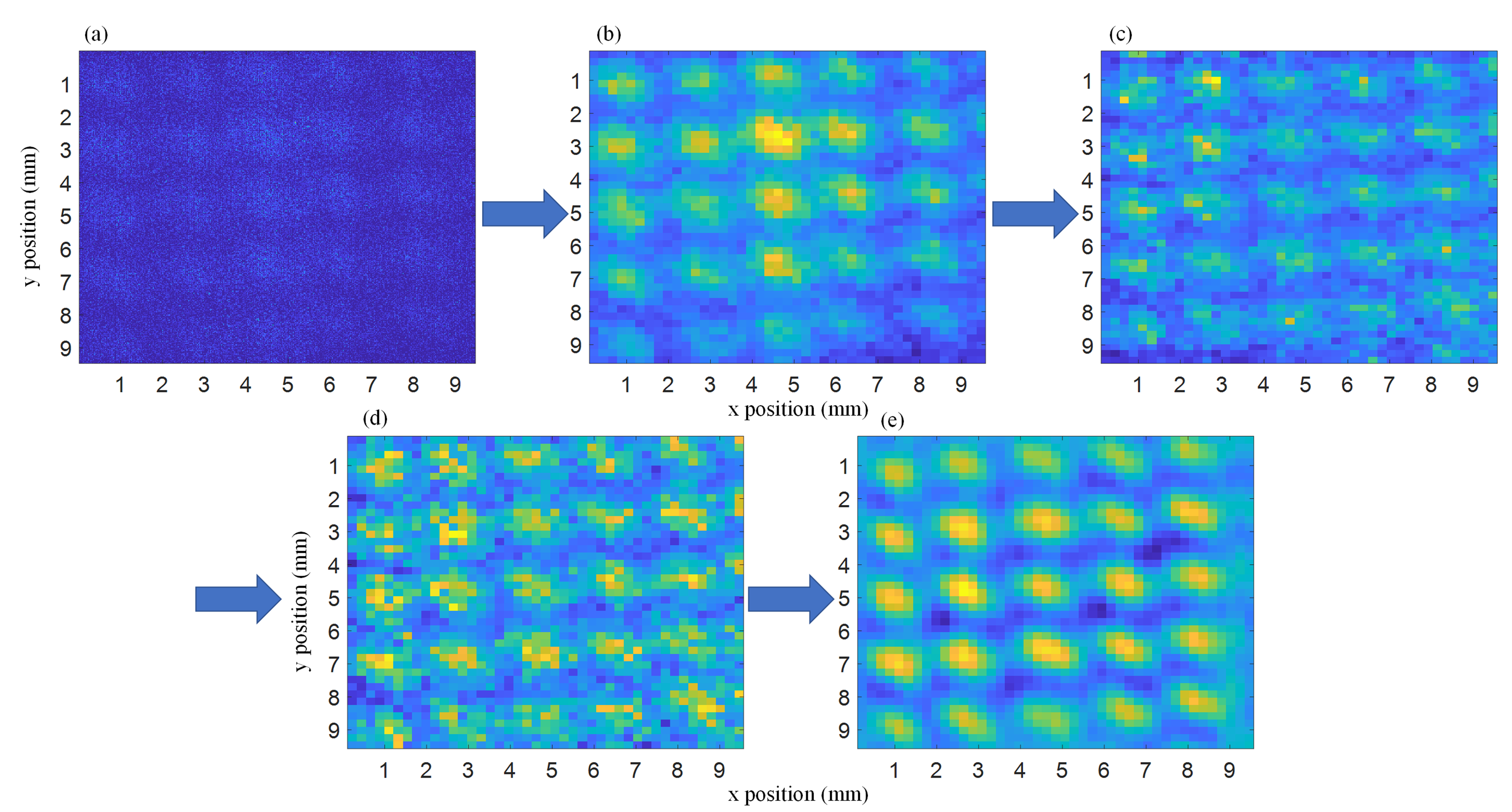}
	\caption{An image of the measured Moire pattern at each step in the data reduction process. The raw image (a) is binned (b) by a factor of $10\times10$ squared pixels. The first normalization is shown in (c), followed by the second normalization (d) by dividing by a quadratic polynomial. Finally the Fourier filter is applied resulting in the last image (e)}
	\label{steps}
\end{figure*}
The four plate interferometer, described in detail in \cite{Heinrich1988,Hasegawa1996,Filipp2005,Geppert2018}, is part of the S18 beamline situated at the high flux reactor of the Institute Laue Langevin (ILL) in Grenoble France \cite{S18data}. Both the interferometer and the monochromator are cut from a perfect silicon crystal. Both utilize the 220 plane. For this experiment a Bragg angle of 30° was used resulting in a wavelength of $1.92$ \AA \ and a bandwidth of $\Delta\lambda/\lambda\approx0.02$. Our 5° prisms were milled out of aluminium (type: EN AW-6060). We estimate a refraction angle of $1.1 \cdot 10^{-7}$ rad or $0.022$ arcseconds. The phase shifters consisted out of a polished 3 mm thick sapphire slab and a 3 mm thick silicon slab. A $2\,\times \,2\,\mathrm{mm^{2}}$ collimator was placed just downstream from the monochromator, roughly 3 m from the detector. The beam expanded to $10\,\times \,10\,\mathrm{mm^{2}}$ over this distance, indicating a maximal divergence of $\sim$ 2 mrad. This is used to calculated the vertical coherence. The rocking full width half maximum is around half a second of arc, which is used for calculating the horizontal coherence. A position sensitive detector, using a scintillator, a 45° mirror and a CCD camera was employed to record the Moire patterns. The resolution of the CCD camera was determined experimentally to be around 22 $\mu$m, however the scintillator limits the maximally achievable resolution to 40 $\mu m$. 
\subsection{Data Processing}
The images shown in this paper were generated using two recorded datasets, the first with prisms inserted in the interferometer and the second without prisms. Each dataset consists of 14 recordings 30 minutes a piece. The raw images shown in figure \ref{raw} show the summed averages over all 14 images.
To obtain the image shown in figure \ref{Results} both raw images (figure \ref{raw}) are binned by a factor of $10\times10$ squared pixels to increase statistics. Next the binned "prisms-in" image is divided by the binned "prisms-out" image to remove artifacts induced by uneven illumination of the detector and spatial phase drifts inherent to the interferometer. After this initial normalization the resulting image still has a slight intensity drift over the vertical (y) direction. This is removed by fitting a quadratic polynomial to the average intensity along the vertical direction and dividing the image by this polynomial. After this step the image is normalized by dividing it by the mean intensity and subtracting one $I_{norm}=\frac{I}{<I>}-1$. Finally a noise reduction scheme is applied to improve the overall signal quality. This is achieved by Fourier transforming the image, removing all content from the FT below a certain noise floor and transforming the modified FT back to real space. Figure \ref{steps} shows the image at each point of the data reduction.
\subsection{Fitting Procedure}
The fit shown in figure \ref{Results} is based on equation (\ref{equation:Intensity}), with a few modifications to take into account perturbations not considered in our simplified theory. By normalizing the data as described above we forfeit the need for a constant offset in the fit function. In addition to account for the dephasing which reduces the contrast towards the edges of the image since the neutrons have to pass through more material, we multiply (\ref{equation:Intensity}) by a Gaussian envelope. Finally we note that the interference pattern in the images indicate that the prisms were not totally orthogonal, as a result the fit function takes on the form
\begin{equation}\label{equation:Fit}
    \begin{aligned}
        &f=e^{\frac{(\Vec{x}-\Vec{\mu})^2}{s^2}}[a_1\cos{(\Vec{\eta}_1\cdot\Vec{x}+\alpha_1')}+a_2\cos{(\Vec{\eta}_2\cdot\Vec{x} +\alpha_2')}\\ &+a_3\cos([\Vec{\eta}_1-\Vec{\eta}_2]\cdot\Vec{x}+\alpha_1'-\alpha_2')]
    \end{aligned}
\end{equation}
Here initial guesses for $\Vec{\eta}_1$, $\Vec{\eta}_2$, $\alpha_1'$, $\alpha_2'$ $a_1$, $a_2$ and $a_3$ are extracted from the Fourier transform of the processed data. An initial guess for $\Vec{\mu}$ is found by determining the expectation value of the squared processed data $<\Vec{x}>=\frac{\int \mathrm{d}x\mathrm{d}y \Vec{x} I^2}{\int \mathrm{d}x\mathrm{d}yI^2}$. Finally the parameter $s^2$ is guessed by calculating the variance of the squared processed data
$\frac{\int \mathrm{d}x\mathrm{d}y |\Vec{x}|^2 I^2}{\int \mathrm{d}x\mathrm{d}yI^2}-<\Vec{x}>^2$
\begin{widetext}
    \subsection{Detailed Calculations}
In this subsection the step by step calculations of $<L_z>$ (for isotropic and anisotropic momentum distributions) and $<L_z^2>$ are shown. Starting with the calculation of $<L_z>$ in cylindrical coordinates
$$<L_z>=-\mathrm{i}\frac{\int \mathrm{d}\mathbf{r} \psi_t^*(\mathbf{r})\frac{\partial}{\partial \phi}\psi(\mathbf{r})}{\int \mathrm{d}\mathbf{r} |\psi_t(\mathbf{r})|^2}$$
with
$$\int \mathrm{d}\mathbf{r} |\psi_t(\mathbf{r})|^2=1+\cos(\Delta\alpha)e^{-\frac{\sigma^2 k_\perp^2}{4}}=N$$
and
$$-\mathrm{i}\frac{\partial}{\partial \phi}\psi_t(\mathbf{r})=\frac{1}{\sqrt{2}}\psi_0[k_\perp\rho \cos(\phi)e^{\mathrm{i}k_\perp\rho \sin(\phi)}-e^{\mathrm{i}\Delta\alpha}k_\perp \rho \sin(\phi) e^{\mathrm{i}k_\perp \rho \cos(\phi)}]$$
hence it follows
$$<L_z>=\frac{1}{2N}\int \mathrm{d}\mathbf{r}|\psi_0|^2[k_\perp\rho\cos(\phi)-k_\perp\rho\sin(\phi)+k_\perp\rho\cos(\phi)e^{-\mathrm{i}\Delta\alpha}e^{\mathrm{i}k_\perp\rho (\sin(\phi)-\cos(\phi))}-k_\perp\rho\sin(\phi)e^{\mathrm{i}\Delta\alpha}e^{\mathrm{i}k_\perp\rho (\cos(\phi)-\sin(\phi))}]$$
which, using $\int_0^{2\pi} \mathrm{d\phi} \cos(\phi)=\int_0^{2\pi} \mathrm{d\phi} \sin(\phi)=0$, simplifies to
$$<L_z>=\frac{1}{2N}\int \mathrm{d}\mathbf{r}k_\perp\rho|\psi_0|^2[\cos(\phi)e^{-\mathrm{i}\Delta\alpha}e^{\mathrm{i}k_\perp\rho (\sin(\phi)-\cos(\phi))}-\sin(\phi)e^{\mathrm{i}\Delta\alpha}e^{\mathrm{i}k_\perp\rho (\cos(\phi)-\sin(\phi))}]$$

$$<L_z>=\frac{1}{2N}\int d\mathbf{r}k_\perp\rho|\psi_0|^2[\cos(\phi)e^{-i\Delta\alpha}e^{\mathrm{i}\sqrt{2}k_\perp\rho\sin(\phi-\pi/4)}-\sin(\phi)e^{\mathrm{i}\Delta\alpha}e^{-i\sqrt{2}k_\perp\rho sin(\phi-\pi/4)}]$$
Then we apply the Jacobi-Anger expansion, $e^{\mathrm{i}z\sin(\phi)}=\sum_\ell J_\ell (z) e^{\mathrm{i}\ell\phi}$ \cite{Abramowitz}, and use that $\int_0^{2\pi}\mathrm{d}\phi e^{\mathrm{i}\ell\phi}=0$ for $\ell\neq0$ This allows us to easily solve the azimuthal integral.
$$<L_z>=\frac{\pi}{2N}\int \mathrm{d}\rho k_\perp\rho^2|\psi_0|^2[J_{-1}(\sqrt{2}k_\perp \rho) e^{-\mathrm{i}\Delta\alpha}e^{\mathrm{i}\frac{\pi}{4}}+J_{1}(\sqrt{2}k_\perp \rho) e^{-\mathrm{i}\Delta\alpha}e^{-\mathrm{i}\frac{\pi}{4}}-$$$$\mathrm{i}J_{-1}(\sqrt{2}k_\perp \rho) e^{\mathrm{i}\Delta\alpha}e^{\mathrm{i}\frac{\pi}{4}}+\mathrm{i}J_{1}(\sqrt{2}k_\perp \rho) e^{\mathrm{i}\Delta\alpha}e^{-\mathrm{i}\frac{\pi}{4}}]$$
Next we use the anti-symmetry of the Bessel function of first order $J_{-1}(z)=-J_{1}(z)$ and begin grouping the exponential/trigonometric terms.
$$<L_z>=-\mathrm{i}\frac{\pi}{N}\int \mathrm{d}\rho k_\perp\rho^2|\psi_0|^2J_1(\sqrt{2}k_\perp\rho)[e^{-\mathrm{i}\Delta\alpha}\sin(\frac{\pi}{4})-e^{\mathrm{i}\Delta\alpha}\cos(\frac{\pi}{4})]$$

$$<L_z>=2\pi\frac{\sin\Delta\alpha}{\sqrt{2}N}\int \mathrm{d}\rho k_\perp\rho^2|\psi_0|^2J_1(\sqrt{2}k_\perp\rho)$$
Which can be rewritten into the form of a standard Hankel transform of first order with known result. This brings us to equation \ref{equation:TotalOAM}
$$<L_z>=\sin(\Delta\alpha)\frac{k_\perp^2\sigma^2}{4N}e^{-\frac{\sigma^2k_\perp^2}{4}}$$
Next we examine the generalized case where the momentum distribution is anisotropic (see equation \ref{equation:trueInput}). This is best done in Cartesian coordinates:
$$L_z\psi_t=-\mathrm{i} (x\frac{\partial}{\partial y}-y\frac{\partial}{\partial x})\psi_t$$

$$=-\frac{\mathrm{i}}{\sqrt{\pi \sigma_x\sigma_y}}e^{-\frac{x^2}{\sigma_x^2}-\frac{y^2}{\sigma_y^2}}(2xy(e^{\mathrm{i}k_\perp y}+e^{\mathrm{i}\Delta\alpha}e^{\mathrm{i}k_\perp x})[\frac{1}{\sigma_x^2}-\frac{1}{\sigma_y^2}]+\mathrm{i}k_\perp x e^{\mathrm{i}k_\perp y}-\mathrm{i}k_\perp y e^{\mathrm{i}\Delta\alpha}e^{\mathrm{i}k_\perp x})$$
Hence it follows

$$<L_z>=\frac{k_\perp}{\pi \sigma_x\sigma_y N} \int \mathrm{d}x \mathrm{d}y e^{-2\frac{x^2}{\sigma_x^2}-2\frac{y^2}{\sigma_y^2}} [xe^{-\mathrm{i}k_\perp(x-y)-\mathrm{i}\Delta\alpha}-y e^{\mathrm{i}k_\perp(x-y)+\mathrm{i}\Delta\alpha}]$$
Where all odd terms have been dropped since their integral is zero. To proceed we use $\mathrm{i}\frac{\partial}{\partial k}e^{-\mathrm{i}k(a+b)}=(a+b)e^{-\mathrm{i}k(a+b)}$ to get
$$<L_z>=\frac{k_\perp}{\pi \sigma_x\sigma_y N} \int \mathrm{d}x \mathrm{d}y e^{-2\frac{x^2}{\sigma_x^2}-2\frac{y^2}{\sigma_y^2}} [\mathrm{i}\frac{\partial}{\partial k_\perp}e^{-\mathrm{i}k_\perp(x-y)-\mathrm{i}\Delta\alpha}-\mathrm{i}\frac{\partial }{\partial k}e^{\mathrm{i}k_\perp(x-y)+\mathrm{i}\Delta\alpha}+ye^{-\mathrm{i}k_\perp(x-y)-\mathrm{i}\Delta\alpha}-x e^{\mathrm{i}k_\perp(x-y)+\mathrm{i}\Delta\alpha}]$$
Note the final term is minus the complex conjugate of our previous expression for $<L_z>$ hence it follows

$$<L_z>=\frac{k_\perp}{\pi \sigma_x\sigma_y N} \int \mathrm{d}x \mathrm{d}y e^{-2\frac{x^2}{\sigma_x^2}-2\frac{y^2}{\sigma_y^2}} [\mathrm{i}\frac{\partial}{\partial k_\perp}e^{-\mathrm{i}k_\perp(x-y)-\mathrm{i}\Delta\alpha}-\mathrm{i}\frac{\partial }{\partial k}e^{\mathrm{i}k_\perp(x-y)+\mathrm{i}\Delta\alpha}]-<L_z>^*$$
and since expectation values must be real we can conclude

$$<L_z>=\frac{\mathrm{i}k_\perp}{2\pi \sigma_x\sigma_y N} \int \mathrm{d}x \mathrm{d}y e^{-2\frac{x^2}{\sigma_x^2}-2\frac{y^2}{\sigma_y^2}} \frac{\partial}{\partial k_\perp}[e^{-\mathrm{i}k_\perp(x-y)-\mathrm{i}\Delta\alpha}-e^{\mathrm{i}k_\perp(x-y)+\mathrm{i}\Delta\alpha}]$$
We may now swap integration and differentiation and realize that we are left with a standard Fourier transform
$$<L_z>=\frac{\mathrm{i}k_\perp}{2\pi \sigma_x\sigma_y N} \frac{\partial}{\partial k_\perp}\int \mathrm{d}x \mathrm{d}y e^{-2\frac{x^2}{\sigma_x^2}-2\frac{y^2}{\sigma_y^2}} [e^{-\mathrm{i}k_\perp(x-y)-\mathrm{i}\Delta\alpha}-e^{\mathrm{i}k_\perp(x-y)+\mathrm{i}\Delta\alpha}]$$
Conducting the transform and grouping the exponential/trigonometric terms leads to
$$<L_z>=\frac{k_\perp}{2N} \frac{\partial}{\partial k_\perp}e^{-\frac{(\sigma_x^2+\sigma_y^2)k^2}{8}} \sin(\Delta\alpha)$$
Finally carrying out the differentiation leads to the result shown in equation \ref{equation:ASOAM}
$$<L_z>=\sin(\Delta\alpha) \frac{k_\perp^2(\sigma_x^2+\sigma_y^2)}{8N} e^{-\frac{k_\perp^2 (\sigma_x^2+\sigma_y^2)}{8}}$$
Finally we calculate the second moment of the OAM distribution, $<L_z^2>$
$$<L_z^2>=-\frac{\int d\mathbf{r} \psi_t^*(\mathbf{r})\frac{\partial^2}{\partial \phi^2}\psi_t(\mathbf{r})}{N}$$

$$\frac{\partial^2}{\partial \phi^2}\psi(\mathbf{r})=-\frac{1}{\sqrt{2}}\psi_0[(k_\perp^2 \rho^2 \sin^2(\phi)+\mathrm{i}k_\perp \rho \cos(\phi)) e^{\mathrm{i}\Delta\alpha}e^{\mathrm{i}k_\perp \rho \cos(\phi)}+ (k_\perp^2\rho^2 \cos^2(\phi)+\mathrm{i}k_\perp\rho \sin(\phi))e^{\mathrm{i}k_\perp\rho \sin(\phi)}]$$
Therefore
$$<L_z^2>=\frac{1}{2N}\int  \mathrm{d}\mathbf{r}  |\psi_0|^2[(k_\perp^2 \rho^2 \sin^2(\phi)+\mathrm{i}k_\perp \rho \cos(\phi))+(k_\perp^2\rho^2 \cos^2(\phi)+\mathrm{i}k_\perp\rho \sin(\phi))+$$$$(k_\perp^2 \rho^2 \sin^2(\phi)+\mathrm{i}k_\perp \rho \cos(\phi))e^{\mathrm{i}\Delta\alpha} e^{\mathrm{i}k_\perp \rho [\cos(\phi)-sin(\phi)]}+(k_\perp^2\rho^2 \cos^2(\phi)+\mathrm{i}k_\perp\rho \sin(\phi))e^{-\mathrm{i}\Delta\alpha}e^{\mathrm{i}k_\perp\rho [\sin(\phi)-\cos(\phi)]}]$$
First we use $\cos(\phi)-\sin(\phi)=-\sqrt{2}\sin(\phi-\frac{\pi}{4})$
$$<L_z^2>=\frac{1}{2N}\int  \mathrm{d}\mathbf{r}  |\psi_0|^2[k_\perp^2 \rho^2 \sin^2(\phi)+k_\perp^2\rho^2 \cos^2(\phi)+$$$$(k_\perp^2 \rho^2 \sin^2(\phi)+\mathrm{i}k_\perp \rho \cos(\phi)) e^{\mathrm{i}\Delta\alpha}e^{-\mathrm{i}\sqrt{2}k_\perp \rho \sin(\phi-\frac{\pi}{4})}+(k_\perp^2\rho^2 \cos^2(\phi)+\mathrm{i}k_\perp\rho \sin(\phi))e^{-\mathrm{i}\Delta\alpha}e^{\mathrm{i}\sqrt{2}k_\perp\rho \sin(\phi-\frac{\pi}{4})}]$$
We simplify the expression by using the identity $\cos^2(\phi)+\sin^2(\phi)=1$
$$<L_z^2>=\frac{1}{2N}\int  \mathrm{d}\mathbf{r}  |\psi_0|^2[k_\perp^2 \rho^2 +(k_\perp^2 \rho^2 \sin^2(\phi)+\mathrm{i}k_\perp \rho \cos(\phi)) e^{\mathrm{i}\Delta\alpha}e^{-\mathrm{i}\sqrt{2}k_\perp \rho \sin(\phi-\frac{\pi}{4})]}+$$$$(k_\perp^2\rho^2 \cos^2(\phi)+\mathrm{i}k_\perp\rho \sin(\phi))e^{-\mathrm{i}\Delta\alpha}e^{\mathrm{i}\sqrt{2}k_\perp\rho \sin(\phi-\frac{\pi}{4})}]$$
We solve the azimuthal integral by using use the Jacobi-Anger expansion again, $e^{\mathrm{i}z\sin(\phi)}=\sum_\ell J_\ell (z) e^{\mathrm{i}\ell\phi}$, and again use that $\int_0^{2\pi}\mathrm{d}\phi e^{\mathrm{i}\ell\phi}=0$ for $\ell\neq0$. Note that the latter identity paired with the trigonometric terms in the previous line filter out all but the $\ell=0$ and $\ell=\pm1$ terms of the Jacobi-Anger expansion.
$$<L_z^2>=\frac{1}{2N}\int \mathrm{d}\rho \rho |\psi_0|^2[2\pi k_\perp^2 \rho^2 +(\pi k_\perp^2 \rho^2 J_0(\sqrt{2}k_\perp \rho)-\sqrt{2}\pi k_\perp \rho J_1(\sqrt{2}k_\perp \rho) )e^{\mathrm{i}\Delta\alpha}+$$$$(\pi k_\perp^2\rho^2 J_0(\sqrt{2}k_\perp \rho)-\sqrt{2}\pi k_\perp\rho J_1(\sqrt{2}k_\perp \rho))e^{-\mathrm{i}\Delta\alpha}]$$
Here we have once again used the asymmetry of the first order Bessel function. Next we group together the trigonometric terms
$$<L_z^2>=\frac{1}{2N}\int \mathrm{d}\rho \rho |\psi_0|^2[2\pi k_\perp^2 \rho^2 +\cos(\Delta\alpha)(2\pi k_\perp^2 \rho^2 J_0(\sqrt{2}k_\perp \rho)-\sqrt{8}\pi k_\perp \rho J_1(\sqrt{2}k_\perp \rho)]$$

Now we attempt to solve the radial integrals

$$<L_z^2>=\frac{1}{\pi \sigma^2 N}\int \mathrm{d}\rho \rho e^{-2\frac{\rho^2}{\sigma^2}}[2\pi k_\perp^2 \rho^2 +\cos(\Delta\alpha)(2\pi k_\perp^2 \rho^2 J_0(\sqrt{2}k_\perp \rho)-\sqrt{8}\pi k_\perp \rho J_1(\sqrt{2}k_\perp \rho)]$$

The first integral seen above:

$$\int_0^\infty \mathrm{d}\rho 2\pi k_\perp^2 \rho^3 e^{-2\frac{\rho^2}{\sigma^2}} $$

can be solved using integration by parts and substitution ($u=\rho^2$ and $\mathrm{d}u=\rho\mathrm{d}\rho$)

$$\int_0^\infty \mathrm{d}u \pi k_\perp^2 u e^{-2\frac{u}{\sigma^2}} = [-\pi k_\perp^2\frac{u\sigma^2}{2}e^{-\frac{-2u}{\sigma^2}}]^\infty_0+\int_0^\infty \mathrm{d}u \pi k_\perp^2\frac{\sigma^2}{2}e^{-2\frac{u}{\sigma^2}}=\frac{\pi k_\perp^2\sigma^4}{4} $$
The next radial integral in $<L_z^2>$ is a Hankel transform with a known result:
$$2\pi\cos(\Delta\alpha)\int_0^\infty \mathrm{d}\rho  k_\perp^2 \rho^3 e^{-2\frac{\rho^2}{\sigma^2}} J_0(\sqrt{2}k_\perp \rho)=\frac{\pi k_\perp^2\sigma^4}{4}\cos(\Delta\alpha)e^{-\frac{\sigma^2k_\perp^2}{4}}(1-\frac{ k_\perp^2\sigma^2}{4})$$
The final integral is the same Hankel transform as for the first moment
$$-\sqrt{8}\pi\cos(\Delta\alpha)\int d\rho e^{-2\frac{\rho^2}{\sigma^2}} k_\perp \rho^2 J_1(\sqrt{2}k_\perp \rho)=-\cos(\Delta\alpha)\frac{\pi k_\perp^2\sigma^4}{4}e^{-\frac{k_\perp^2\sigma^2}{4}}$$

Hence we find

$$<L_z^2>=\frac{k_\perp^2\sigma^2}{4N}+\cos(\alpha)\frac{ k_\perp^2\sigma^2}{4N}e^{-\frac{\sigma^2k_\perp^2}{4}}(1-\frac{ k_\perp^2\sigma^2}{4})-\cos(\alpha)\frac{ k_\perp^2\sigma^2}{4N}e^{-\frac{k_\perp^2\sigma^2}{4}}$$

$$<L_z^2>=\frac{k_\perp^2\sigma^2}{4N}-\cos(\alpha)\frac{ k_\perp^4\sigma^4}{16N}e^{-\frac{\sigma^2k_\perp^2}{4}}$$
\end{widetext}
\section{Data Availability}
The data that support the findings of this study are available via \cite{S18data}.
\section{References}
\bibliographystyle{unsrt}
\bibliography{SpinOrbitBibliography}
\section{Acknowledgement}
This work was funded by the Austrian Science Fund (FWF), Project No. P34239.
\section{Author Contributions}
N.G. H.L. and S.S. conceived the experiment; N.G. H.L. and A.B. carried out the experiment; N.G. analyzed the data; N.G. wrote the paper with contributions from all authors.
\section{Competing Interests}
The Authors declare no Competing Financial or Non-Financial Interests
\end{document}